\documentclass{anabs}
\usepackage{graphicx}
\usepackage{times}
\Volume{1-2}             
\Year{2003}              
\Month{02}               
\Pagespan{000}{000}      

\begin{document}
\lhead[\thepage]{A.N. Pradas, J. \& Kerp, J.: The 3-D composition of the Galactic interstellar medium. The hot phases and X-ray absorbing material.}
\rhead[Astron. Nachr./AN~{\bf 324} (2003) 1/2]{\thepage}
\headnote{Astron. Nachr./AN {\bf 324} (2003) 1/2, 000--000}

\title{The 3-D composition of the Galactic interstellar medium. The hot phases and X-ray absorbing material.}

\author{J. Pradas \& J. Kerp}
\institute{Radioastronomisches Institut der Universit\"at Bonn, Auf dem H\"ugel 71, 53121 Bonn, Germany}

\correspondence{jpradas@astro.uni-bonn.de}

\maketitle

\section{Introduction}
A detailed investigation of the soft X-ray background (SXRB) - correlation analysis of the ROSAT all-sky survey (Snowden et al. 1995) with the Leiden/Dwingeloo \ion{H}{i} 21\,cm-line survey(Hartmann \& Burton 1997) - leads to a consistent model of the whole SXRB (Pradas, Kerp \& Kalberla, these proceedings)

The high precision of the data allows to identify coherent structures which deviate from the best fit SXRB model. Here, we study these regions and associate them either with high (HVC) and intermediate-velocity clouds (IVC) or with irregularities in the shape of the Local Hot Bubble (LHB).

\section{Analysis and results}

In order to estimate the deviation of the SXRB model to the observational data, we calculate the following value for every pixel of $12\arcmin \times 12\arcmin$ size:

\begin{equation}
\label{deviation}
\delta = \frac{I_{\mathrm{observed}}-I_{\mathrm{modelled}}}{\sigma_{\mathrm{observation}}}
\end{equation}

\begin{figure}
{\includegraphics[scale=0.23]{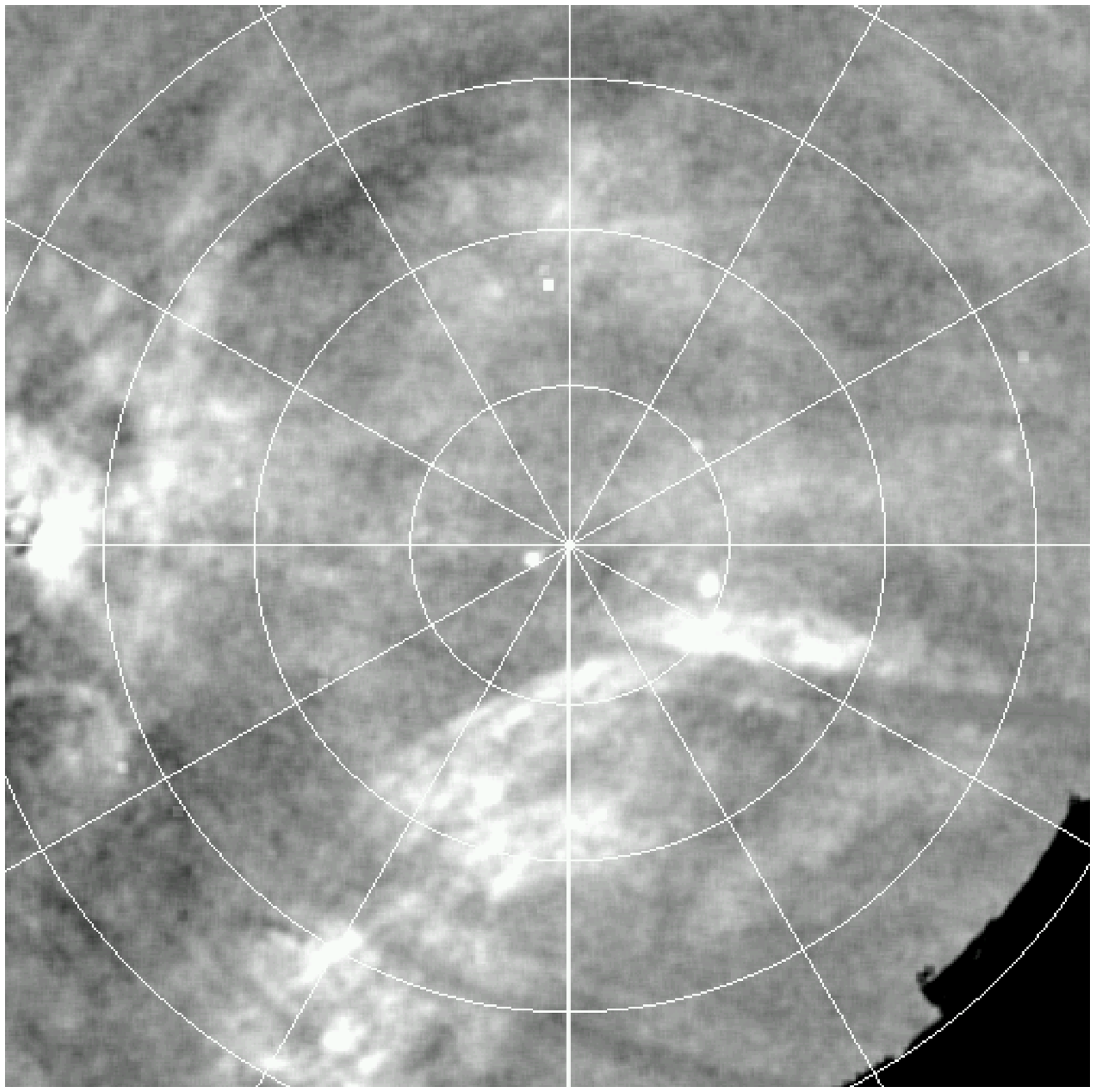}}
{\includegraphics[scale=0.23]{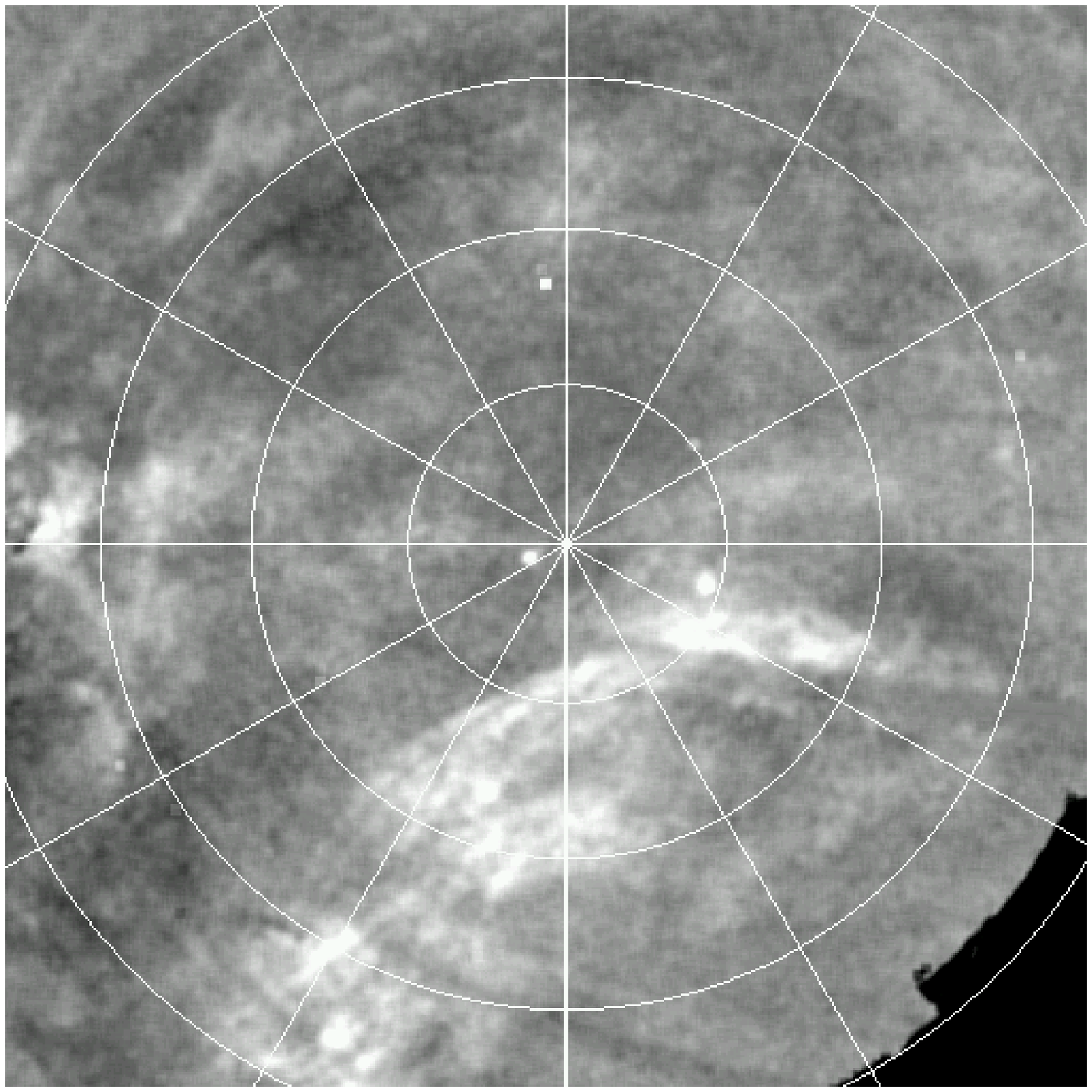}}
\caption{{\bf left:} Deviation image of the ROSAT R2 band model without optimisation for the absorber distribution. Cut levels are $-3\sigma$ and $3\sigma$. The field is centred at the north galactic pole. The step in galactic latitude is $15\degr$ and in galactic longitude is $30\degr$ with $l=0\degr$ directed to the bottom. {\bf right:} The same as before but with optimisation of the absorber distribution. The prominent feature at the bottom is the North Polar Spur.}
\label{devim}
\end{figure}

This statistical variable $\delta$ ideally follows a N(0,1) distribution, and  automatically rejects pixels where the quality of the observational data is low and therefore not relevant for the investigation of the SXRB on large scales. In a first approach, the model is consistent with 80\% of the whole sky in {\em all} seven ROSAT energy bands (see Figure \ref{devim}).
Pixels with big deviations (above $3\sigma$) form coherent structures that are not expected for a N(0,1) distribution. We focus on these areas, which can be interpreted as follows:

\begin{itemize}
\item BRIGHT: Model too faint. Excess of absorber in the transport equation, additional X-ray sources {\em or} lack of intensity in the modelled LHB.
\item DARK: Model too bright. Lack of absorber in the transport equation {\em or} excess of intensity in the modelled LHB. 
\end{itemize}

With this approach we can interpret the deviation images as a 3-D tool to determine the location of neutral gas clouds. 
We vary the locations of the different absorber components relative to the galactic X-ray halo (in front of or beyond it) to find the best fit distribution. This procedure leads to a significantly better agreement between model and observations if features like HVC Complex C and Complex M (Wakker \& van Woerden 1997) are supposed to be located beyond the galactic X-ray halo and, thus, not absorbing the galactic halo emission.

Even with this best fit model, the distribution of $\delta$ still forms coherent structures. These are generally of a smaller size and lower intensity (below $2\sigma$) than the deviations described above. We associate these residual discrepancies with irregularities in the LHB. We conclude that the variation of the intensity of the LHB is a smooth function of galactic coordinates.

\begin{acknowledgements}
  The authors like to thank the Deutsches Zentrum f\"ur Luft- und
  Raumfahrt for financial support under grant No. 50 OR 0103.
\end{acknowledgements}

\end{document}